\documentclass{ws-procs9x6}            
\begin{document}
\title{On nonlinear fractional maps: \\
Nonlinear maps with power-law memory}

\author{M. Edelman$^*$}

\address{Dept. of Physics, Stern College at Yeshiva University,\\
245 Lexington Ave, New York, NY 10016, USA; \\
Courant Institute of Mathematical Sciences, New York University,\\
251 Mercer St., New York, NY 10012, USA;\\
Department of Mathematics, BCC, CUNY,\\
2155 University Avenue, Bronx, New York 10453, USA\\
$^*$E-mail: edelman@cims.nyu.edu\\
http://www.yu.edu/faculty/pages/Edelman-Mark}

\begin{abstract}
This article is a short review of the recent results on 
properties of nonlinear fractional maps which are maps with 
power- or asymptotically power-law memory. These maps demonstrate the new
type of attractors - cascade of bifurcations type trajectories, power-law
convergence/divergence of trajectories, period doubling bifurcations with changes in 
the memory parameter, intersection of trajectories, and overlapping of
attractors.
In the limit of small time steps these maps converge to nonlinear
fractional differential equations. 
   
\end{abstract}


\bodymatter

\section{Introduction}

The subject of this short review is related to two areas of science, fractional
calculus and systems with memory, which have an important
overlapping: fractional dynamical systems (systems which are described by
differential equations with fractional time derivatives (FDE) or, in the case of
discrete systems, fractional difference equations) are systems with
power-law memory.
Recent interest in fractional calculus is stimulated by  its 
wide applicability to description and modeling of various natural, social,
and engineering systems. Popularity of fractional calculus and FDE can be 
demonstrated, for example, by the large number of
books on the subject published over the last few years (see, e.g., 
Ref.~\citenum{MainardiBook2010,TarasovBook2011,FrDyn2011,ControlBook2011,Nour}).

Applications of fractional calculus in physics include:
{\bf a.} Fractional transport and fractal properties of time. 
In Hamiltonian systems and billiards
\cite{ZasBook2008} fractal structure of phase space and
stickiness of trajectories in time imply description of transport by
the fractional (fractional time and space derivatives) Fokker-Plank-Kolmogorov
equation;
{\bf b.} Systems of oscillators with long range interaction
\cite{TarasovBook2011,LuoBook2010,LZ2006,ZET2007}; 
{\bf c.} Dielectric Materials  \cite{TDia2009};
{\bf d.} Viscoelastic materials 
\cite{MainardiBook2010,CM1971,Visc4,Visc5,Wineman2007,NLVisc2}; etc.

Biological systems are systems in which memory appears naturally and
is a significant factor defining their evolution. Hierarchy of building blocks  
of biological systems, from
individual neurons and proteins to tissues of individual organs,
demonstrate power-law memory,  $\sim
t^{-\beta}$, with $-1<\beta<1$ (see multiple references in
Ref.~\citenum{MarkBook2014}). 
Power-law memory has been demonstrated in  
processing of external stimuli by individual neurons \cite{Lund2,Lund1},
adaptation of biological systems at levels ranging from 
single ion channels up to human psychophysics   \cite{Adaptation3}, and human memory
(forgetting - the accuracy on memory tasks decays 
as a power law \cite{Kahana,
Adaptation1}) and learning (the reduction in reaction times that comes with
practice is a power function of the number of training trials \cite{Anderson}).
Note, that fractional maps
corresponding to fractional differential and difference equations of the order
$0<\alpha<2$  are maps with  power-law memory in which the power is
$-\beta=\alpha-1$, where $-1<\beta<1$ \cite{DNC}.

The subjects of population biology are various kinds of populations,
which in many cases are not memoryless. The basic discrete model in
population biology is the logistic map \cite{May}, which is essentially 
nonlinear.
This suggests the importance of the investigation of basic properties 
of nonlinear maps with memory.
  
In what follows, in Section \ref{memory} we describe the main results on equivalence of systems with power-law memory (maps and integral Volterra equations of the second kind) to fractional differential/fractional difference equations, in Section \ref{maps} we present various forms of fractional/fractional difference maps, three figures in Section \ref{properties} illustrate main properties of systems with power-law memory, and in Section \ref{conclusion} we discuss applications and perspectives of the research of systems with power-law memory.

\section{Maps with Power-Law Memory}
\label{memory}

The form of maps with memory which allows detailed investigation still being 
quite general is an one-dimensional  map 
with long-term memory \cite{MM1,MM2,MM6,MM5,MM3,MM7}
\begin{equation}
x_{n+1}=\sum^{n}_{k=0}V_{\alpha}(n,k)G_K(x_k),
\label{LTM}
\end{equation} 
where $x_k$ is a value of a model's (physical) variable at the time $t_k$,  
$n,k\in  \mathbb{Z}$, $n,k \ge 0$, $V_{\alpha}(n,k)$ and 
$\alpha \in  \mathbb{R}$ 
characterize memory effects, and $K$ is a nonlinearity parameter. In the case of
power-law memory it can be written as
\begin{equation}
x_{n}=\sum^{n-1}_{k=0}(n-k)^{\alpha-1} G_K(x_k,h),
\label{LTMPL}
\end{equation}
where $h$ is a constant time step between $t_n$ and $t_{n+1}$, 
$n \in  \mathbb{N}$, and $\alpha>0$. This map is equivalent to the fractional difference
equation \cite{Chaos2015}  
\begin{equation}
\sum^{n}_{k=0}
(-1)^k 
\left( \begin{array}{c}
\alpha \\ k
\end{array} \right)
x_{n-k}=
(-1)^n
\left( \begin{array}{c}
\alpha \\ n
\end{array} \right)
x_0 +
\sum^{n-1}_{k=0}G_K(x_{n-k-1},h) A(\alpha-1,k),
\label{nstep}
\end{equation}
where \cite{Podlubny,Samko}
\begin{equation}
\left( \begin{array}{c}
\alpha \\ n
\end{array} \right)
=
\frac{\alpha(\alpha-1)...(\alpha-n+1)}{n!}=\frac{\Gamma(\alpha+1)}
{\Gamma(n+1)\Gamma(\alpha-n+1)} 
\label{FracCombinations}
\end{equation}
and the Eulerian numbers with fractional order parameters are defined as 
\cite{BH}
\begin{equation}
A(\alpha,k)=
\sum^{k}_{j=0}
(-1)^j 
\left( \begin{array}{c}
\alpha+1 \\ j
\end{array} \right)
(k+1-j)^{\alpha}.
\label{FracEulerian}
\end{equation} 
A more general result can be formulated as the following theorem  
\cite{Chaos2015}

\begin{theorem}
Any long term memory map 
\begin{equation}
x_{n}=\sum^{\lceil \alpha \rceil - 1}_{k=1}\frac{c_k}{\Gamma(\alpha-k+1)}
(nh)^{\alpha-k}  +\sum^{n-1}_{k=0}(n-k)^{\alpha-1} G_K(x_k,h),
\label{FrLTMPLNN}
\end{equation}
where $\alpha \in  \mathbb{R}$, is equivalent to the map 
\begin{eqnarray}
&&\sum^{n}_{k=0}
(-1)^k 
\left( \begin{array}{c}
\alpha \\ k
\end{array} \right)
x_{n-k}-\sum^{\lceil \alpha \rceil - 1}_{i=1}\frac{c_ih^{\alpha-i}}
{\Gamma(\alpha-i+1)}
\sum^{i-1}_{k=0}
(-1)^k 
\left( \begin{array}{c}
i-1 \\ k
\end{array} \right)
A(\alpha-i,n-k-1) \nonumber \\
&&=(-1)^n
\left( \begin{array}{c}
\alpha \\ n
\end{array} \right)
x_0 +
\sum^{n-1}_{k=0}G_K(x_{n-k-1},h) A(\alpha-1,k).
\label{nstepNN}
\end{eqnarray}
\label{The7}
\end{theorem}
Assuming 
\begin{equation}
G_K(x,h)=\frac{1}{\Gamma (\alpha)}h^{\alpha}G_K(x), \ \ x=x(t), \ \  x_k=x(t_k), \ \ t_k=a+kh, \ \ nh=t-a,
\label{GLDef}
\end{equation}
where $G_K(x)$ is continuous  (then $x(t) \in C^m$) and  $0 \le k \le n$,
in the limit  $h \rightarrow 0+$ for $\alpha>0$ 
Theorem \ref{The7} yields the following result \cite{Chaos2015}
\begin{theorem}
For $\alpha \in  \mathbb{R}$, $\alpha > 0$ 
The Volterra integral equation of the second kind
\begin{equation}
x(t)=
\sum^{\lceil \alpha \rceil}_{k=1}\frac{c_k}{\Gamma(\alpha-k+1)}(t-a)^{\alpha-k}+
\frac{1}{\Gamma (\alpha)}
\int^{t}_{a}\frac{G_K(x(\tau))d\tau}{(t-\tau)^{1-\alpha}}, \  \ \ (t>a),
\label{VoltReal}
\end{equation}
where $G_K(x(\tau))$ is a continuous on $x \in [x_{min}(\tau), x_{max}(\tau)]$,
$\tau\in[a,t]$ function is equivalent to the fractional differential equation 
\begin{equation}
_aD^{\alpha}_tx(t)=G_K(x(t)),
\label{Theorem3b}
\end{equation}
where the derivative on the left side is the Gr$\ddot{u}$nvald-Letnikov
fractional derivative,
with the initial conditions 
\begin{equation}
c_k=(_aD^{\alpha-k}_tx)(t=a+), \  \  \ k=1,...,\lceil \alpha \rceil. 
\label{IC1}
\end{equation}
\label{The5}
\end{theorem}
Similar theorems are also proven for the differential equations with 
Caputo and Riemann-Liouville fractional derivatives\cite{Kilbas}. 
These results were used\cite{Kilbas} to prove the existence and uniqueness 
of solutions of fractional differential equations. Maps with power-law 
memory Eq.~(\ref{FrLTMPLNN}) can be calculated for any  
functions $G_K(x,h)$ and any $c_k$, which implies the existence and uniqueness of
solutions of the corresponding fractional difference equations 
(Eq.~(\ref{nstepNN})). 
Equivalence of fractional differential equations and 
Volterra integral equations of the second kind was used to derive
fractional maps\cite{DNC,TZmap,Tmap1,Tmap2,Chaos2013}.

\section{Fractional Maps}
\label{maps}

Results presented in the previous section suggest that maps with 
real power-law memory Eq.~(\ref{FrLTMPLNN}) can be used to study
general properties of fractional dynamical systems (see Section \ref{properties}). In this section we present the other three forms of maps with power- asymptotically power-law memory used to investigate general properties of systems with memory. 

Equations similar to 
Eq.~(\ref{FrLTMPLNN}) are derived from fractional differential  
equations describing systems with periodic kicks 
\cite{DNC,TZmap,Tmap1,Tmap2,Chaos2013} 
\begin{equation}
\frac{d^{\alpha}x}{dt^{\alpha}}+G_K(x(t- \Delta T)) \sum^{\infty}_{k=-\infty} \delta \Bigl(\frac{t}{T}-(k+\varepsilon)
\Bigr)=0,   
\label{UM1D2Ddif}
\end{equation} 
where $\varepsilon > \Delta > 0$,  $\alpha \in \mathbb{R}$, $\alpha>0$, in
the limit $\varepsilon  \rightarrow 0$ with the initial conditions
corresponding to the type of fractional derivative we are going to
use. The case $\alpha =2$, $\Delta = 0$, and $G_K(x)=KG(x)$
corresponds to the equation used to derive the universal map in regular 
dynamics (see, e.g., Ref.~\citenum{ZasBook2008}).
This is why the equations obtained by integrating Eq.(\ref{UM1D2Ddif}):
\begin{equation}
x^{(s)}_{n+1}= \sum^{N-s-1}_{k=0}\frac{x^{(k+s)}_0}{k!}(n+1)^{k} 
-\frac{1}{\Gamma(\alpha-s)}\sum^{n}_{k=0} G_K(x_k) (n-k+1)^{\alpha-s-1},
\label{FrCMapx}
\end{equation} 
where $s=0,1,...,N-1$,  $0 \le N-1 < \alpha \le N$ are
called the universal Caputo map and the equations
{\setlength\arraycolsep{0.5pt}
\begin{eqnarray}
&&x_{n+1}=  \sum^{N-1}_{k=1}\frac{c_k}{\Gamma(\alpha-k+1)}(n+1)^{\alpha -k} \nonumber \\  
&&-\frac{1}{\Gamma(\alpha)}\sum^{n}_{k=0} G_K(x_k) (n-k+1)^{\alpha-1}, 
\label{FrRLMapx} \\
&&p^s_{n+1}= \sum^{N-s-1}_{k=1}\frac{c_k}{(N-s-1-k)!} (n+1)^{N-s -1-k}
\nonumber \\  
&&-\frac{1}{(N-s-2)!}\sum^{n}_{k=0} G_K(x_k) (n-k+1)^{N-s-2},
\label{FrRLMapp} 
\end{eqnarray} }
where $s=0,1,...N-2$ 
are called the universal Riemann-Liouville map.

For the discrete systems the fractional difference Caputo universal map 
\cite{Chaos2014,DNC2015,Baleanu} (of order $\alpha$) can be written as 
\begin{equation} 
x_{n+1} =   \sum^{m-1}_{k=0}\frac{\Delta^{k}x(0)}{k!}(n+1)^{(k)} 
-\frac{1}{\Gamma(\alpha)}  
\sum^{n+1-m}_{s=0}(n-s-m+\alpha)^{(\alpha-1)} 
G_K(x_{s+m-1}), 
\label{FalFacMap}
\end{equation}
where $\Delta^{k}x(0)=c_k$ for  $k=0, 1, ..., m-1$, 
are the initial conditions and $m=\lceil \alpha \rceil$. The falling
factorial function is defined as 
\begin{equation}
t^{(\alpha)} =\frac{\Gamma(t+1)}{\Gamma(t+1-\alpha)}
\label{FrFac}
\end{equation}
and the map is an asymptotically power-law memory map because
\begin{equation}
\lim_{s \rightarrow
  \infty}\frac{\Gamma(s+\alpha)}{\Gamma(s+1)s^{\alpha-1}}=1,  
\ \ \ \alpha \in  \mathbb{R}.
\label{GammaLimit}
\end{equation}

\section{Fractional/Fractional Difference Standard and Logistic $\alpha$-Families of Maps}
\label{properties}

Eqs.~(\ref{FrCMapx})-(\ref{FalFacMap}) 
with $G_K(x)=Ksin(x)$, which
for $\alpha=2$ yield the regular standard map, are called the fractional 
standard $\alpha$-families of maps (Caputo,
Riemann-Liouville, and difference Caputo, correspondingly). 
Eqs.~(\ref{FrCMapx})-(\ref{FalFacMap}) with 
$G_K(x)=x-Kx(1-x)$, which for $\alpha=1$ 
yield the regular logistic map, 
are called the fractional logistic $\alpha$-families of maps (Caputo,
Riemann-Liouville, and Caputo difference, correspondingly). Initial
investigation of the general properties of fractional dynamical systems 
(systems with power- or asymptotically power-law memory) was performed
using the fractional standard and logistic $\alpha$-families of maps
\cite{MarkBook2014,DNC,Chaos2015,Chaos2013,Chaos2014,DNC2015,ETFSM,Universality,TEdisFM,myFSM,Taieb}. 

\begin{figure}
\begin{center}
\includegraphics[width=4.5in]{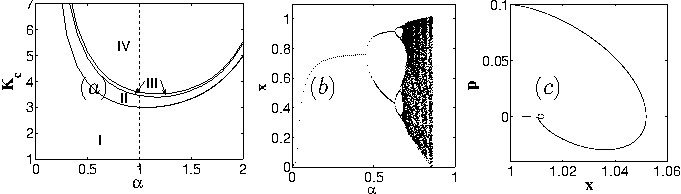}
\end{center}
\caption{Bifurcations and self-intersections of trajectories:
(a) $\alpha-K$ diagrams for the logistic Caputo $\alpha$-family of maps. The
fixed point is stable in the area I, the $T=2$ sink is
stable in the area II, the area III is the area of
the stable $T=2^n$ ($n>1$, $n \in \mathbb{N}$) sinks and CBTT, 
and the area IV is the area of chaos; 
(b) bifurcations with the change in the memory parameter in figure (a) for 
$K=4.2$;
(c) a self-intersecting phase space trajectory of the fractional Caputo 
Duffing equation $_0^CD^{1.5}_t x(t)=x(1-x^2)$, $t\in [0,40]$ 
with the initial conditions $x(0)=1$
and $p_0=dx/dt(0)=0.1$.
}
\label{Fig1}
\end{figure}

\begin{figure}
\begin{center}
\includegraphics[width=4.5in]{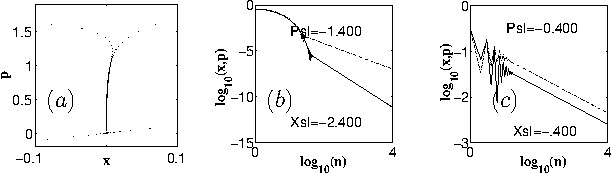}
\end{center}
\caption{Power-law convergence of trajectories:
(a) two converging to the origin trajectories of the standard Riemann-Liouville
$\alpha$-family of maps with $\alpha=1.4$ and $K=2$. The upper slow 
converging trajectory, $x_n \sim n^{-\alpha}$ and 
$p_n \sim n^{1-\alpha}$, starts from the chaotic area ($x_0=0$, $p_0=5.5$). 
The lower fast converging trajectory, $x_n \sim n^{-1-\alpha}$ and 
$p_n \sim n^{-\alpha}$, starts from the attractor's basin of attraction
($x_0=0$, $p_0=0.3$); 
(b) $x$ and $p$ time dependence for the fast converging trajectory in (a);
(c) $x$ and $p$ time dependence for trajectories of the standard 
Caputo $\alpha$-family of maps with $\alpha=1.4$ and $K=2$. All
converging trajectories follow the power law $x_n \sim n^{1-\alpha}$ and 
$p_n \sim n^{1-\alpha}$.
}
\label{Fig2}
\end{figure}

\begin{figure}
\begin{center}
\includegraphics[width=4.5in]{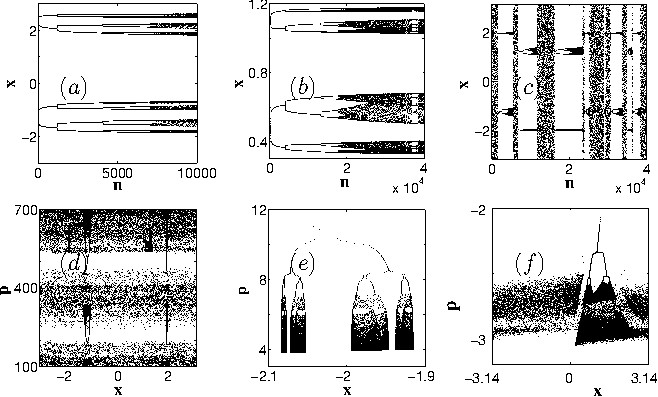}
\end{center}
\caption{Attracting cascade of bifurcations type trajectories:
(a) a single attracting trajectory of the standard Caputo fractional difference
$\alpha$-family of maps with $\alpha=0.1$ and $K=2.41$; 
(b)  a single attracting trajectory of the logistic Caputo 
$\alpha$-family of maps with $\alpha=0.1$ and $K=22.65$;
(c)  a single intermittent CBTT of the standard Riemann-Liouville
$\alpha$-family of maps with $\alpha=1.65$ and $K=4.5$;
(d) the phase space for the case (c);
(e) a half of the attractor in the phase space of the standard 
Riemann-Liouville $\alpha$-family of maps with $\alpha=1.1$ and $K=3.5$;
(f) CBTT (initial conditions $x_0=0$ and $p_0=-1.8855$) overlapping with a 
chaotic attractor (initial conditions $x_0=0$ and $p_0=-2.5135$) of the  
standard Caputo $\alpha$-family of maps with $\alpha=1.02$ and $K=4.5$.
}
\label{Fig3}
\end{figure}

In spite of some differences, bifurcations with changes in 
the memory parameter, intersection of trajectories and overlapping of
chaotic attractors, power-law convergence/divergence of trajectories,
and the new type of attractors - cascade of bifurcations type trajectories (CBTT) were demonstrated in all $\alpha$-families of maps. 
They are illustrated in Figs.~\ref{Fig1}-\ref{Fig3}.
Unlike in integer maps, in maps with power- asymptotically power-law memory periodic attractors (sinks)
exist only in asymptotic sense. If a system with memory begins its evolution
from an attracting point then it leaves this point and may end its
evolution at the same or at a different attracting point.

As in the case of one-dimensional nonlinear maps, for $0<\alpha<2$ all
members of all $\alpha$-families (map equations with fixed $\alpha$) demonstrate period doubling cascades of
bifurcations scenarios of transition to chaos with the change in the
nonlinearity parameter $K$. Existence of self-similarity and the corresponding
constants (similar to the Feigenbaum constants) is not yet
investigated. The corresponding $x-K$ bifurcation diagrams depend on $\alpha$
and it results in 2-dimensional bifurcation diagrams. An example of the
$\alpha-K$ bifurcation diagram for the logistic Caputo $\alpha$-family of
maps is presented in Fig.~\ref{Fig1}~(a). It is obvious from this 
two-dimensional diagram that in addition to $x-K$ bifurcation diagrams
for the fixed $\alpha$ there exist $x-\alpha$ bifurcation diagrams for the
fixed $K$. An example of such a diagram is presented in
Fig.~\ref{Fig1}~(b). 

In systems with long-term memory the next value of the system's variable $x$
depends on the whole history of the system's evolution. It is clear that in
such systems individual trajectories may intersect or self-intersect and 
chaotic attractors may overlap. An example of a self-intersecting
trajectory in the fractional Caputo Duffing oscillator $_0^CD^{1.5}_t
x(t)=x(1-x^2)$, which is a continuous system with power-law memory, is
presented in Fig.~\ref{Fig1}~(c). Two overlapping chaotic attractors in the
standard Caputo family of maps, one of which is a CBTT, are presented in 
Fig.~\ref{Fig3}~(f). 

The power-law convergence of trajectories to attracting points ($1<\alpha<2$)
is demonstrated in Figs.~\ref{Fig2} (b) and (c). In Caputo maps 
the convergence is slower than in the corresponding Riemann-Liouville maps.
Phase space of fractional maps may contain attracting points with 
their basins of attraction and a chaotic
sea from which two neighboring points may converge to different
attractors. Unlike the situation in integer maps, the rate of convergence 
in fractional maps depends on the initial conditions. Trajectories
which start from basins of attraction converge faster than those that
start from a chaotic sea. An example of such two differently converging 
trajectories in the standard
Riemann-Liouville map is given in Fig.~\ref{Fig2}~(a).

One of the remarkable features of maps with power- asymptotically power-law
memory is existence of attracting cascade of bifurcations type
trajectories. Convergence to sinks of the period $T=2^n$ with
$n>0$ may follow various scenarios. Trajectories may converge directly 
to a $T=2^n$ sink following a power law; 
they may initially converge to a sink of lower periodicity and then, through a few period doubling bifurcations 
converge to a $T=2^n$ sink; there is also a possibility
(this way trajectories converge to attractors in Caputo $\alpha$-family of
maps \cite{Chaos2013} with $1<\alpha<2$) 
that initially trajectories converge to a higher
periodicity sink and then converge to a $T=2^n$ sink
through a series of inverse bifurcations.  
Near the border with chaos (e.g., upper border of the area  
III in Fig.~\ref{Fig1}~(a)) the
trajectories converge to CBTT. CBTT look like cascades of bifurcations in
regular dynamics but bifurcations occur on a single attracting
trajectory, Figs.~\ref{Fig3} (a), (b), and (e), without any changes in maps'
memory ($\alpha$) or nonlinearity ($K$) parameters. Possible hidden 
symmetries in maps with power-law memory which lead to CBTT are not found yet. 
Another interesting phenomenon is intermittent CBTT
(see Figs.~\ref{Fig3} (c) and (d)) in which a trajectory may start as a
chaotic trajectory, then converge to a periodic one, then turn back 
into a chaotic trajectory through a period doubling cascade of bifurcations, and this 
chain of transformations will repeat again and again. Existence of cascade
of bifurcations type trajectories in phase spaces of systems with
$1<\alpha<2$ is illustrated in Figs.~\ref{Fig3} (e) and (f).

\section{Conclusion}
\label{conclusion}

The results presented in this review are relevant to all natural and social 
systems with power-law memory. Let us mention a few possible areas of applications. 
In social sciences, the intermittent cascade of bifurcations evolution can be seen 
in evolution of systems on every level, starting from varying opinions of individuals 
and up to the evolution of a society as a whole. The history is a chain of repeating 
periods of democracy, chaos, and dictatorship. 

The fact that in systems with power-law memory bifurcations depend on both 
nonlinearity and memory parameters may be particular interesting in medical 
application. As we mentioned in the introduction, almost each body organ's 
tissues possess nonlinear viscoelastic properties and are nonlinear systems 
with power-law memory. There is even a Maple application 
{\it Nonlinear Viscoelastic Behaviour of Brain Tissue}
(http://www.maplesoft.com/applications/view.aspx?SID=153923\&view=html).
If a disease is associated with changes in a nonlinearity parameter (whatsoever it is), 
then cure can be associated with changes in memory parameter (which could be associated 
with corrections in the nervous system). There are five types of surgical procedures 
to treat epilepsy.
Each of them imposes changes in brain and correspondingly in memory. As the brain 
tissue is a nonlinear system with memory, one of the possible explanations of the 
success in the surgical treatment is that these changes in memory compensate for 
the abnormal value of the brain tissue nonlinearity parameter.    

Definitely, all above mentioned systems are very complicated with many parameters, 
but in situations when one of the parameters is the most important to control the 
state of a system, nonlinear equations (maps or differential equations) with 
power-law memory can be used to model the system in order to explain and predict 
its evolution.  

One of the basic tools in the analysis of nonlinear one-dimensional maps,
the return map, makes no sense in the case of maps with memory, because the next 
value of a system variable $x_{n+1}$ depends on all previous values of this variable. 
The analysis of bifurcation diagrams in this case is quite complicated because 
individual trajectories may bifurcate and the rate of convergence to attracting 
points follows the power law $n^{-\alpha}$, which for small $\alpha$ typical for 
biological applications is very slow. The problem of finding some general form of 
self-similarity in maps with power-law memory is an open and important problem.

CBTT are found in discrete maps and to find them in continues systems described by 
fractional differential equations is another interesting problem. There are examples 
of continuous fractional chaotic systems of dimensions less than three but more than 
two. To prove impossibility of chaos in fractional continuous systems of lower 
dimensions or to find the counterexample is also an important open problem in the 
analysis of nonlinear systems with power-law memory.

\section*{Acknowledgments}

The author acknowledges support from the Stern College at
Yeshiva University.  
The author expresses his gratitude to the organizers of the 
Chaos Complexity and Transport (CCT15) conference in Marseilles for their kind
invitation  
and to  G. Ben Arous and M. J. Shelley
for the opportunity to complete this work at the Courant Institute.

\end{document}